\documentstyle[12pt]{article}

\newcommand{\bce}{\begin{center}}
\newcommand{\ece}{\end{center}}
\newcommand{\beq}{\begin{equation}}
\newcommand{\eeq}{\end{equation}}
\newcommand{\bea}{\vspace{0.25cm}\begin{eqnarray}}
\newcommand{\eea}{\end{eqnarray}}

\newcommand{\ba}{\begin{array}}
\newcommand{\ea}{\end{array}}

\newcommand{\doublespace}{
    \renewcommand{\baselinestretch}{1.6}\large\normalsize}

\def\lsim{\mathrel{\rlap{\lower4pt\hbox{\hskip1pt$\sim$}}
    \raise1pt\hbox{$<$}}}	  %less than or approx. symbol
\def\gsim{\mathrel{\rlap{\lower4pt\hbox{\hskip1pt$\sim$}}
    \raise1pt\hbox{$>$}}}	  %greater than or approx. symbol

  \advance\oddsidemargin  by -1in
  \advance\evensidemargin by -1in
  \advance\topmargin      by -0.5in
\def\lsim{\mathrel{\rlap{\lower4pt\hbox{\hskip1pt$\sim$}}
    \raise1pt\hbox{$<$}}}         %less than or approx. symbol
\def\gsim{\mathrel{\rlap{\lower4pt\hbox{\hskip1pt$\sim$}}
    \raise1pt\hbox{$>$}}}         %greater than or approx. symbol

\def\beq{\begin{equation}}
\def\endeq{\end{equation}}
\def\arr{\begin{eqnarray}}
\def\endarr{\end{eqnarray}}
\makeindex

 \doublespace

\begin{document}

%\phantom{.}{\bf \Large \hspace{10.0cm} SISSA-............ \\  }
\phantom{.}{\bf \Large \hspace{10.0cm} KFA-IKP(Th)-1994-12\\ }
\phantom{.}\hspace{10.9cm}7 March 1994\\
%}

\begin{center}
{\bf \huge On determination of the large-${1\over x}$ \\ gluon
distribution at HERA
\vspace{1.0cm}\\}
{\bf \LARGE
N.N.~Nikolaev$^{a,b}$,
and B.G.~Zakharov$^{b,c}$ \vspace{1.0cm} \\}
{\it
$^{a}$IKP(Theorie), KFA J{\"u}lich, 5170 J{\"u}lich, Germany
\medskip\\
$^{b}$L. D. Landau Institute for Theoretical Physics, GSP-1,
117940, \\
ul. Kosygina 2, Moscow V-334, Russia \\
$^{c}$Interdisciplinary Laboratory for Advanced Studies (ILAS)\\
Miramare, I-34014 Trieste, Italy \vspace{1.0cm}\\
}
{\LARGE
Abstract}\\

\end{center}
We discuss corrections to the Leading-Log$Q^{2}$ relationships
between the gluon density $g(x,Q^{2})$ and
$F_{L}(x,Q^{2}),\,\partial F_{T}(x,Q^{2})
/\partial \log Q^{2}$ in the HERA range of large ${1\over x}$.
We find that the above quantities probe the gluon density
$g(x,Q_{T,L}^{2})$ at $Q_{T,L}^{2}=C_{T,L}Q^{2}$, with
the $Q^{2}$-rescaling factors
$C_{T}\approx 2.2$ and $C_{L}\approx 0.42$. The possibility
of treating charm as an active flavour is critically re-examined.
 \bigskip\\

\begin{center}
E-mail: kph154@zam001.zam.kfa-juelich.de
\end{center}

%\doublespace
\pagebreak

%-----------------

%                 Section  1

%-----------------

\section{ Introduction.}

%-----------------
At large ${1\over x}$\,, the gluon desnity $g(x,Q^{2})$
is much higher than the
density of charged partons $q(x,Q^{2}),\bar{q}(x,Q^{2})$
(hereafter $x$ is the
Bjorken variable and $Q^{2}$ is the virtuality of the photon) and
the photoabsorption will be
dominated by interaction with the nucleon of the lightcone
$q\bar{q}$  Fock states of the photon via the exchange by gluons
(Fig.~1) or, alternatively, by the photon-gluon fusion
$\gamma^{*}g \rightarrow q\bar{q}$. Consequently, the longitudinal
structure function $F_{L}(x,Q^{2})$ and the slope of the
transverse structure function $F_{T}(x,Q^{2})=2xF_{1}(x,Q^{2})$
become direct probes of the gluon structure function
$G(x,Q^{2})=xg(x,Q^{2})$:
\beq
F_{L}(x,Q^{2})=
{\alpha_{S}(Q^{2}) \over 3\pi} \sum e_{f}^{2}
G(\xi_{L}x,Q^{2}) \, ,
\label{eq:1.1}
\endeq
\beq
{\partial F_{T}(x,Q^{2}) \over \partial \log(Q^{2})}=
{\alpha_{S}(Q^{2}) \over 3\pi} \sum e_{f}^{2}
G(\xi_{T}x,Q^{2})
\label{eq:1.2}
\endeq
In Eqs.~(\ref{eq:1.1},\ref{eq:1.2}) $\alpha_{S}(Q^{2})$ is the
running strong coupling, $e_{f}$ is the quark charge in units
of the electron charge and the summation goes over the active
flavours $f$. The suggestion of $F_{L}$ as a probe (partonometer)
of the gluon
density is due to Dokshitzer [1], Eq.~(\ref{eq:1.1}) was
elaborated in [2]. Eq.~(\ref{eq:1.2}) readily follows from
formulas (6.37) and (6.34) of the Roberts'
textbook [3] (see also [4]).
Both equations were derived in the
Leading-Log$Q^{2}$ approximation (LLQA),
the $x$-rescaling factors $\xi_{T,L} \approx 2$ [2-4].
The emergence of the gluon-dominated scaling violations at large
${1\over x}$ was
clearly demonstrated in the recent QCD-evolution analysis of
the NMC structure functions [5]. The obvious advantage of
Eqs.~(\ref{eq:1.1},\ref{eq:1.2}) is that one does not need
solving the coupled QCD-evolution equations for the gluon and
(anti)quark densities.

As a byproduct of our analysis [6] of determination of the BFKL
pomerons intercept from $F_{L}(x,Q^{2}),
\partial F_{T}(x,Q^{2})/\partial \log Q^{2}$ we have noticed
the potentially important corrections to the LLQA. One obvious
issue is which $Q^{2}$ is sufficiently large for charm to be
treated as an active flavor, because for the $(u,d,s)$ active
flavours $\sum e_{f}^{2}={2\over 3}$ compared to
$\sum e_{f}^{2} ={10\over 9}$ if charm also were an active flavour.
Our analysis [6] suggests that
it is premature to speak of $N_{f}=4$ active flavours unless
$Q^{2}\gsim (100-200)$GeV$^{2}$. In [6] we also noticed that
$F_{L}(x,Q^{2})$ and $\partial F_{T}(x,Q^{2})/\partial \log Q^{2}$
probe the gluon structure function $G(x,Q_{T,L}^{2})$ at different
values of $Q_{T,L}^{2}$, which are both different from $Q^{2}$.
Because of importance of determination of the gluon structure
function, which is a fundamental quantity in the QCD parton
model, in this paper we present an update of
Eqs.~(\ref{eq:1.1},\ref{eq:1.2}) which are valid also in the
BFKL (Balitzkii-Fadin-Kuraev-Lipatov [7]) regime, i.e., beyond
the LLQA.

At large ${1\over x}$, deep inelastic scattering can be
viewed as an interaction with the target nucleon of the lightcone
$q\bar{q}$  Fock states of the photon via the exchange by
gluons (Fig.1). The principal quantities are the probability
densities $|\Psi_{T,L}(z,r)|^{2}$ for the $q\bar{q}$ Fock states
with the transverse size $\vec{r}$ and the fraction $z$ of
photon's lightcome momentum carried by the (anti)quark, and
$\sigma(x,r)$ -  the total cross section of interaction
of the $q\bar{q}$ colour dipole of transverse size $r$ with
the nucleon target [8,9]. This dipole cross section satisfies the
generalized BFKL equation derived in [9,10] and is related to
the differential density of gluons by the equation [9,11,12]
\beq
\sigma(x,r)={\pi \over 3 }\alpha_{S}(r)r^{2}\int
{ d^{2}\vec{k}\over k^{2}}\cdot
{4[1-\exp(i\vec{k}\,\vec{r})]\over k^{2}r^{2} }
{\partial G(x_{g},k^{2})  \over \partial \log k^{2}} \, .
\label{eq:1.3}
\endeq
The wave functions of the $q\bar{q}$ Fock states of the
(T) transverse and (L) longitudinal photon were
derived in [8] and read
\beq
\vert\Psi_{T}(z,r)\vert^{2}=
\sum e_{f}^{2}|\Psi_{T}^{(f\bar{f})}(z,r)|^{2}=
{6\alpha_{em} \over (2\pi)^{2}}
\sum_{1}^{N_{f}}e_{f}^{2}
\{[z^{2}+(1-z)^{2}]\varepsilon^{2}K_{1}(\varepsilon r)^{2}+
m_{f}^{2}K_{0}(\varepsilon r)^{2}\}\,\,,
\label{eq:1.4}
\endeq
\beq
\vert\Psi_{L}(z,r)\vert^{2}=
\sum e_{f}^{2}|\Psi_{T}^{(f\bar{f})}(z,r)|^{2}=
{6\alpha_{em} \over (2\pi)^{2}}
\sum_{1}^{N_{f}}4e_{f}^{2}\,\,
Q^{2}\,z^{2}(1-z)^{2}K_{0}(\varepsilon r)^{2}\,\,,
\label{eq:1.5}
\endeq
where $K_{\nu}(x)$ are the modified Bessel functions,
$\varepsilon^{2}=z(1-z)Q^{2}+m_{f}^{2}$ and
$m_{f}$ is the quark mass. The resulting
photoabsorption cross sections are equal to ([8], see also [13])
\arr
\sigma_{T}(\gamma^{*}N,x,Q^{2})=
\int_{0}^{1} dz\int d^{2}\vec{r}\,\,
\vert\Psi_{T}(z,r)\vert^{2}\sigma(x,r)=
{2\alpha_{em}\over \pi}\sum_{f} e_{f}^{2}
\int_{0}^{1} dz \int {d^{2}\vec{k}\over k^{4}}~~~~~~~~
\nonumber\\
\int {d^{2}\vec{\kappa} \over  \vec{\kappa}^{2}+\varepsilon^{2}}
\left\{ {[z^{2}+(1-z)^{2}]\vec{k}^{2} +m_{f}^{2} \over
\vec{k}^{2}+\varepsilon^{2}}-
{[z^{2}+(1-z)^{2}]\vec{k}(\vec{k}+\vec{\kappa})+m_{f}^{2} \over
(\vec{k}+\vec{\kappa})^{2}+\varepsilon^{2}}\right\}
{\partial G(x_{g},k^{2}) \over \partial \log k^{2}}
\alpha_{S}(q^{2})\, ,~~~~
\label{eq:1.6}
\endarr
\arr
\sigma_{L}(\gamma^{*}N,x,Q^{2})=
\int_{0}^{1} dz\int d^{2}\vec{r}\,\,
\vert\Psi_{L}(z,r)\vert^{2}\sigma(x,r)=
{2\alpha_{em}\over \pi}\sum_{f} e_{f}^{2}
\int_{0}^{1} dz\, 4Q^{2}z^{2}(1-z)^{2}
\int {d^{2}\vec{k}\over k^{4}}  \nonumber\\
\int {d^{2}\vec{\kappa} \over \vec{\kappa}^{2}+\varepsilon^{2}}
\left\{ {1\over  \vec{k}^{2}+\varepsilon^{2}}-
{1 \over \vec{k}(\vec{k}+\vec{\kappa})+\varepsilon^{2}}\right\}
{\partial G(x_{g},k^{2}) \over \partial \log k^{2}}
\alpha_{S}(q^{2}) \, , ~~~~~~~~~~~~~~~
\label{eq:1.7}
\endarr
Here the running coupling $\alpha_{S}(q^{2})$
enters the integrand at the largest
relevant virtuality,
\beq
q^{2}={\rm max }\{\varepsilon^{2}+\kappa^{2},k^{2}\}\, ,
\label{eq:1.8}
\endeq
and the density of gluons enters at
\beq
x_{g}=x(1+{M_{t}^{2} \over Q^{2}}) \,
\label{eq:1.9}
\endeq
where $M_{t}$ is the transverse mass of the produced $q\bar{q}$
pair in the photon-gluon fusion $\gamma^{*}g\rightarrow q\bar{q}$:
\beq
M_{t}^{2} = {m_{f}^{2}+\vec{\kappa}^{2} \over 1-z}+
{m_{f}^{2}+(\vec{\kappa}+\vec{k})^{2} \over z}   \, .
\label{eq:1.10}
\endeq
The flavour and $Q^{2}$ dependence of structure functions
is concentrated in wave functions (\ref{eq:1.4},\ref{eq:1.5}),
whereas the dipole cross section $\sigma(x,r)$ (the differential
gluon density $\partial G(x,k^{2}) / \partial \log k^{2}$
in the momentum representation) is universal for all
flavours. The important virtue of the $(\vec{r},z)$ representation
in (\ref{eq:1.6},\ref{eq:1.7}) is the factorization of integrands as
$|\Psi_{T,L}(z,r)|^{2}\sigma(x,r)$, which
corresponds to the {\sl exact}
diagonalization of the diffraction scattering matrix in the
$(\vec{r},z)$-representation. Furthermore, the above dipole-cross
section representation (\ref{eq:1.6},\ref{eq:1.7}) and wave functions
(\ref{eq:1.4},\ref{eq:1.5}) are valid also in the BFKL regime,
i.e., beyond the LLQA, and allow  an easy incorporation of the
unitarity (absorption) corrections at large ${1\over x}$ [9,11],
with allowance for which Eq.~(\ref{eq:1.3}) must rather be regarded
as an operational definition of the gluon density beyond the LLQA.
The $x$ (energy) dependence of the
dipole cross section $\sigma(x,r)$ comes from the higher
$q\bar{q}g_{1}...g_{n}$ Fock states of the photon, i.e., from
the QCD evolution effects, described at large ${1\over x}$
by the generalized BFKL equation [9,10]. The structure functions
are given by the familar equation
$
F_{T,L}(x,Q^{2})=(Q^{2}/4\pi \alpha_{em})
\sigma_{T,L}(x,Q^{2}).
$
We advocate using $F_{L}$ and $F_{T}=2xF_{1}$, which have simpler
theoretical interpretation compared to $F_{2}=F_{T}+F_{L}$ which
mixes interactions of the transverse and longitudinal photons.
For the sake of completeness, we notice that,
in the Born approximation  the differential gluon density is
related to the two-body formfactor of the nucleon
$\langle N|\exp(i\vec{k}_{1}\vec{r}_{1}+i\vec{k}_{2}\vec{r}_{2})
|N\rangle$ by the equation [12]
\beq
{\partial G(x,k^{2})\over \partial \log k^{2}}=
{4\over \pi}\alpha_{S}(k^{2})(1-
\langle N|\exp(i\vec{k}(\vec{r}_{1}-\vec{r}_{2})
|N\rangle \, ,
\label{eq:1.11}
\endeq
and the limiting form of Eqs.~(\ref{eq:1.6},\ref{eq:1.7})
derived in [8] is obtained.

The leading contribution comes from values of $M_{t}^{2}\sim Q^{2}$,
so that $x_{g}\sim 2x$. Strictly speaking, the
$x$-rescaling factors can not be determined within the
Leading-Log${1\over x}$ approximation. From the practical
point of view, when analysing $F_{L}$ and $\partial F_{T}/\partial
\log Q^{2}$, it is sufficient to use $x_{g}=\xi_{T,L}x$ with
$\xi_{T,L}\approx 2$ as determined in [2-4]. In this communication
we concentrate on corrections to the LLQA.

%--------------------------------------

%                           Section 2

%----------------------

\section{Active flavours and the onset of LLQA}

The ratio $\sigma(x,r)/r^{2}$ is a smooth function of $r$.
Similarly, $\partial G(x,k^{2})/\partial \log k^{2}$ is a
smooth function of $k^{2}$. Consequently,
it is convenient to use the representations
\beq
F_{T}(x,Q^{2}) = {1\over \pi^{3}} \int {dr^{2}\over r^{2}}
{\sigma(x,r)\over r^{2}}
\sum e_{f}^{2}\Phi_{T}^{(f\bar{f})}(Q^{2},r^{2})\, ,
\label{eq:2.1}
\endeq
\arr
F_{L}(x,Q^{2}) = {1\over \pi^{3}} \int {dr^{2}\over r^{2}}
{\sigma(x,r)\over r^{2}}
\sum e_{f}^{2}W_{L}^{(f\bar{f})}(Q^{2},r^{2})\nonumber\\=
{\alpha_{S}(Q^{2}) \over 3\pi} \sum e_{f}^{2}
\int {dk^{2}\over k^{2}}\Theta_{L}^{(f\bar{f})}(Q^{2},k^{2})
{dG(\xi_{L}x,k^{2})\over d\log k^{2}}\, ,
\label{eq:2.2}
\endarr
\arr
{\partial F_{T}(x,Q^{2})\over \partial \log Q^{2}} =
{1\over \pi^{3}} \int {dr^{2}\over r^{2}}
{\sigma(x,r)\over r^{2}}
\sum e_{f}^{2}W_{T}^{(f\bar{f})}(Q^{2},r^{2})\\
 =
{\alpha_{S}(Q^{2}) \over 3\pi} \sum e_{f}^{2}
\int {dk^{2}\over k^{2}}\Theta_{T}^{(f\bar{f})}(Q^{2},k^{2})
{dG(\xi_{T}x,k^{2})\over d\log k^{2}} \, ,
\label{eq:2.3}
\endarr
where the weight functions $\Phi_{T}^{(f\bar{f})}$ and
$W_{T,L}^{(f\bar{f})}$ are defined by
\beq
\Phi_{T}^{(f\bar{f})}(Q^{2},r^{2}) =
(\pi^{2}/4\alpha_{em})\int_{0}^{1} dz \, Q^{2}r^{4}
|\Psi_{T}^{(f\bar{f})}(z,r)|^{2}\, ,
\label{eq:2.4}
\endeq
\beq
W_{L}^{(f\bar{f})}(Q^{2},r^{2}) =
(\pi^{2}/4\alpha_{em})\int_{0}^{1} dz \, Q^{2}r^{4}
|\Psi_{L}^{(f\bar{f})}(z,r)|^{2}\,,
\label{eq:2.5}
\endeq
\beq
W_{T}^{(f\bar{f})}(Q^{2},r^{2}) =
{\partial \Phi_{T}^{(f\bar{f})}(Q^{2},r^{2})\over
\partial \log Q^{2}}\, ,
\label{eq:2.6}
\endeq
and the kernels $\Theta_{T,L}$ are given by
\beq
\Theta_{T,L}(Q^{2},k^{2})=\int {dr^{2}\over r^{2}}
{\alpha_{S}(q^{2}) \over
\alpha_{S}(Q^{2})}
{4[1-J_{0}(kr)]\over (kr)^{2}}
W_{T,L}(Q^{2},r^{2}) \, .
\label{eq:2.7}
\endeq
Here $J_{0}(x)$ is the Bessel function and
the running coupling $\alpha_{S}(q^{2})$
enters at the largest relevant virtuality:
 $q^{2}={\rm max}\{ k^{2},C^{2}/r^{2}\}$, where
$C\approx 1.5$ ensures the numerically similar results of
calculations in the $(r,z)$ and the momentum representations [8].
Since the BFKL equation is only known to the leading order,
and the differences between the leading-order and next-to-leading
order at large $Q^{2}$ are marginal [3], below we use the
one-loop strong coupling $\alpha_{S}(k^{2})=
4\pi/\beta_{0}\log(k^{2}/\Lambda^{2})$ with $\Lambda =0.3$GeV.
Here $\beta_{0}=11-{2\over 3}N_{f}$, and in
the numerical estimates we impose the infrared freezing
$\alpha_{S}(k^{2})\leq \alpha_{S}^{(fr)}=0.8$.

The conventional LLQA corresponds to
\beq
\Phi_{T}^{(f\bar{f})}(Q^{2},r^{2})=\theta(Q^{2}-{1\over r^{2}})\, ,
\label{eq:2.8}
\endeq
\beq
\Theta_{T,L}^{(f\bar{f})}(Q^{2},k^{2})=\theta(Q^{2}-k^{2})
\label{eq:2.9}
\endeq
and to the factoring out
$\alpha_{S}(q^{2})\approx \alpha_{S}(Q^{2})$
from the integral (\ref{eq:2.7}). A derivation of the LLQA formulae
(\ref{eq:1.1},\ref{eq:1.2}) is based upon precisely these
approximations. Our analysis of approximation (\ref{eq:2.8}) in
[6] revealed a very slow onset of LLQA for the charmed quarks.
Below we analyse in more detail
an accuracy of Eq.~(\ref{eq:2.9}) and of the LLQA relations
(\ref{eq:1.1},\ref{eq:1.2}).

Our results for $\Theta_{T,L}^{(f\bar{f})}(Q^{2},k^{2})$
are shown in Fig.~2. We find a very strong departure from the
LLQA Eq.~(\ref{eq:2.9}): i) the kernels $\Theta_{T,L}$ have
a very broad diffuse edge, ii) the position of the diffuse
edge is shifted compared to the naive expectation $k^{2}=Q^{2}$,
iii) the kernels $\Theta_{T,L}$ flatten at large $Q^{2}$, but
the height of the plateau is different from unity, iv) the onset
of LLQA for charm is very slow. Discussion
of these effects is particulary simple in terms of the
representation (\ref{eq:2.7}).

In order to facilitate the
further discussion, we remind the salient features of weight
functions $W_{T,L}^{(f\bar{f})}(Q^{2},r^{2})$ [6]. Firstly, at
asymptotically large $Q^{2}$,
\beq
\int {dr^{2}\over r^{2}}\, W_{T,L}^{(f\bar{f})}(Q^{2},r^{2}) =1\,.
\label{eq:2.10}
\endeq
Secondly, $W_{T,L}^{(f\bar{f})}(Q^{2},r^{2})$ are peaked at
$r^2 =B_{T,L}/Q^{2}$, where
\arr
B_{T}\approx 2.3\, ,\\ B_{L}\approx 11 \, .
\label{eq:2.11}
\endarr
Therefore, $F_{L}(x,Q^{2})$ and $\partial F_{T}(x,Q^{2})/\partial
\log Q^{2}$ probe $\sigma(x,r)$ at $r^{2}=B_{L,T}/Q^{2}$.
Notice a substantial departure from the naive LLQA expectation
of $B_{T,L}\sim 1$. The width of these peaks is quite
broad, $\Delta \log(Q^{2}r^{2})\approx 3$ [6].
Furthermore, the function $f(x)=4[1-J_{0}(x)]
/x^{2}$, shown in Fig.~3, is similar to the step-function, but
has a very broad diffuse edge. The position of the edge
corresponds to an effective step-function
approximation $f(x)=\theta(A_{\sigma}-x)$
with $A_{\sigma}\approx 10$, and leads to a dramatic deviation
from the naive LLQA estimate $Q^{2}\approx 1/r^{2}$ in the
small-$r$ limit of Eq.~(\ref{eq:1.3}):
\beq
\sigma(x,r)\approx {\pi^{2}\over 3} r^{2}\alpha_{S}(r)
G(x,Q^{2}\approx {A_{\sigma}\over r^{2}}) \, .
\label{eq:2.12}
\endeq
Now we discuss the results for $\Theta_{T,L}(Q^{2},k^{2})$ in the
light of these observations.

The diffuse edge of kernels $\Theta_{T,L}$ originates from the
two factors: a relatively large width of the weight functions
$W_{T,L}(Q^{2},r^{2})$, and a broad diffuse edge of
$f(kr)=4[1-J_{0}(kr)]/(kr)^{2}$.

The kernels $\Theta_{T,L}$ flatten at $k^{2}\ll Q^{2}$, but the
height of the plateau $H_{T,L}^{(f\bar{f})}(Q^{2})$
is significantly different from unity. At small $Q^{2}$ the
height of the plateau is smaller than unity, which signals
the sub-LLQA for the considered flavour. This effect is
particularly important for heavy flavours (charm, bottom).
For instance, for the charmed
quarks $H_{T,L}^{(c\bar{c})}$ only very slowly rises
with $Q^{2}$ reaching only $\sim .5$ at $Q^{2}=30$GeV$^{2}$,
so that charm is an only marginally active flavor and for the
charm conrtribution to the structure function the LLQA
is badly broken unless $Q^{2} \gsim (100-200)$GeV$^{2}$.
For the $b$-quarks the height of the plateau is $\approx 0.3$
at $Q^{2}=120$GeV$^{2}$ and $\approx 0.7$ at $Q^{2}=
480$GeV$^{2}$. The onset of the LLQA for heavy flavours
is particularly slow in the case of the longitudinal structure
function, because merely by gauge invariance the longitudinal
cross section is suppressed, $\sigma_{L}/\sigma_{T}\propto
Q^{2}/4m_{f}^{2}$ at $Q^{2}\lsim 4m_{f}^{2}$
(for instance, see [14]).

On the other hand, the excess over unity comes from
$\alpha_{S}(q^{2})/\alpha_{S}(Q^{2}) > 1$ in the integrand of
Eq.~(\ref{eq:2.7}). Of course, at the asymptotically large
$Q^{2}$, the height of the plateau $H_{T,L}^{(f\bar{f})}(Q^{2})$
tends to unity, because here the LLQA becomes accurate and
one would have replaced $\alpha_{S}(q^{2})$ in the integrand by
$\alpha_{S}({1\over r^{2}}) \approx \alpha_{S}(Q^{2})$. Closer
inspection of Eq.~(\ref{eq:2.7}) shows that, at the moderately
large $Q^{2}$, besides the width of $W_{T,L}(Q^{2},r^{2})$ and the
diffuse edge of $f(x)$, a large contributor to $\alpha_{S}(q^{2})
>\alpha_{S}(Q^{2})$ is the large value of $A_{\sigma}$.
In the region of $Q^{2}$ of the practical interest for the
HERA experiments, the excess of $H_{T,L}^{(f\bar{f})}(Q^{2})$
over unity is particularly large for the light quarks.

At sufficiently high $Q^{2} \gsim 4m_{f}^{2}$, the position of
the diffuse edge of $\Theta_{T,L}(Q^{2},k^{2})$ in the natural
variable $k^{2}/Q^{2}$ is approximately flavour-independent.
Notice, that the diffuse edge
is definitely shifted towards positive
$\log(k^{2}/Q^{2}) =\log(C_{T}) \sim 1$ for the slope of the
transverse structure function, and for the longitudinal
structure function the position of the edge is shifted towards
negative $\log(k^{2}/Q^{2})=\log(C_{L})\sim -1$. We find
strong departure from the LLQA assumption $C_{T,L}=1$.

%------------------------------------------------------

%                        Section  3

%-------------------
\section{Measuring the gluon distribution}

The differential gluon structure function
 $\partial G(x,Q^{2})/\partial \log Q^{2}$
is a slow and smooth function of $\log Q^{2}$. For this reason,
we can
quantify the shift of the diffuse
edge and the height of the plateau of
global kernels $\Theta_{T,L}(Q^{2},k^{2})$ for $N_{f}=5$ flavours
$(u,d,s,c,b)$ in terms of the effective step-function
parameterization
\beq
 \Theta_{T,L}(Q^{2},k^{2})={9\over 11}\sum_{f=1}^{5} e_{f}^{2}
\Theta_{T,L}^{(f\bar{f})}(Q^{2},k^{2})  ={9\over 11}
H_{T,L}(Q^{2})\theta(C_{T,L}(Q^{2})Q^{2}-k^{2})\, .
\label{3.1}
\endeq
The so-determined
height $H_{T,L}(Q^{2})$ of the effective step-function and the
$Q^{2}$-rescaling factor $C_{T,L}(Q^{2})$ are shown in Fig.~4.
They enter the modified relationships (\ref{eq:1.1},\ref{eq:1.2})
as follows:
\beq
F_{L}(x,Q^{2})=
{\alpha_{S}(Q^{2}) \over 3\pi} {11\over 9} H_{L}(Q^{2})
G(\xi_{L}x,C_{L}(Q^{2})Q^{2})
\label{eq:3.2}
\endeq
\beq
{\partial F_{T}(x,Q^{2}) \over \partial \log(Q^{2})}=
{\alpha_{S}(Q^{2}) \over 3\pi} {11\over 9} H_{T}(Q^{2})
G(\xi_{T}x,C_{T}(Q^{2})Q^{2})
\label{eq:3.3}
\endeq
The viable approximations for the kinematical range of the HERA
experiments are $C_{T}(Q^{2})=2.2$ and $C_{L}(Q^{2})=0.42$.
Because of variations of the factor $\alpha_{S}(q^{2})/
\alpha_{S}(Q^{2})$ in the integrand of (\ref{eq:2.7}),
these results for $C_{T,L}$ slightly differ from, but are
close to, the estimate
$C_{T,L}(Q^{2}) \sim  B_{T,L}/A_{\sigma}$.

In the case of the slope of the
transverse structure function, there is a
significant
numerical cancellation of effects of the suppressed, sub-LLQA
contribution of heavy flavours, and of the enhanced height of
the plateau for light flavours. Because of this numerical
conspiracy, despite the charm not being an active flavour
at all, we find $H_{T}(Q^{2}>10$GeV$^{2})=1$ to better than
10 per cent accuracy in the kinematical range of HERA.
For this reason, the recent estimate [15] of
the gluon structure function from the H1 data on the scaling
violation, using the the $N_{f}=4$ LLQA formula (\ref{eq:1.2}),
must be regarded as numerically reliable.
For the longitudinal structure function
this cancellation is less complete. The residual departure of
$H_{L}(Q^{2})$ from unity is a monotonic function of $Q^{2}$
and remains quite substantial in the kinematical range of
HERA .

The accuracy of relations (\ref{eq:3.2},\ref{eq:3.3}) can be
checked computing first the structure functions for certain
parametrization of the gluon density, and then comparing the
input gluon density with the ouptut from
(\ref{eq:3.2},\ref{eq:3.3}). We have performed such a test
for the gluon densities [11,12] and [16], both of which give a
good description of the HERA data on the proton structure
function [17]. For the derivative of the transverse structure
function, the input/output comparison suggests that the
accuracy of the above procedure is as good as 5 per cent
at $Q^{2}>10$GeV$^{2}$ in the whole HERA range of $x$.
For the longitudinal structure function, the input/output
agreement is better than 10 per cent at $x>10^{-4}$ and
$Q^{2}>10$GeV$^{2}$, but gets worse at smaller $x$ and
not so large $Q^{2}$. At $x=10^{5}$ the 10 per cent agreement
only holds at $Q^{2}>30$GeV$^{2}$.
These estimates of the accuracy of relations $(\ref{eq:3.2},
\ref{eq:3.3})$ can easily be improved with the advent of
high accuracy data on the gluon structure functions from the
HERA experiments. Small corrections from the $z-r^{2}$
correlations in the wavefunctions (\ref{eq:1.4}.\ref{eq:1.5})
and/or the $x_{g}-(\vec{k},\vec{\kappa})$ correlations implied
by Eqs.~(\ref{eq:1.9},\ref{eq:1.10}), can also be easily
included should the accuracy of the data require that.

%------------------------------

%                      Section 5

%-----------------

\section{Conclusions}

We re-examined determination of the gluon structure function
$G(x,Q^{2})$
from the longitudinal structure function $F_{L}(x,Q^{2})$ and
from the slope $\partial F_{T}(x,Q^{2}) /\partial \log Q^{2}$
of the transverse structure function,  and derived new relationships
Eq.~(\ref{eq:3.2}) and Eq.~(\ref{eq:3.3}). In the
range of $Q^{2}$ of the interest for HERA experiments, charm is
only marginally active flavour. None the less, because of
the numerical conspiracy of corrections to LLQA for the light and
heavy flavour contributions, the overall normalization factor
$H_{T}(Q^{2}) \approx 1$ as if all the $N_{f}=5$ flavours
$(u,d,s,c,b)$ were active. Here the major effect is the
$Q^{2}$-rescaling $C_{T}\approx 2.2$. We find that the onset of
LLQA for the longitudinal structure function is much slower
than for the transverse structure function. Numerical experiments
suggest that our relationships (\ref{eq:3.2},\ref{eq:3.3}) have
$\lsim 10\%$ accuracy.
\medskip\\
{\bf Acknowledgements:\smallskip\\}
Thanks are due to F.Close, J.Forshaw and R.Roberts for discussions
on the longitudinal structure function. One of the authors (BGZ)
is grateful to J.Speth for the hospitality at IKP, KFA J\"ulich,
where this work started, and to S.Fantoni for the hospitality at
ILAS, Trieste.

\pagebreak

\pagebreak
{\bf \Large Figure captions}
\begin{itemize}
\item[Fig.1]
 - Leading QCD subprocesses at large ${1\over x}$.

\item[Fig.2]
 - The kernels $\Theta_{T,L}(Q^{2},k^{2})$ for the light
quarks $(u,d)$, the charmed quark and the global kernel
for $N_{f}=5$ flavours $(u,d,s,c,b)$. The dot-dashed,
dashed, solid, double-dot-dashed and dotted curves are for
$Q^{2}=0.75,\, 2.5,\,8.5,\,30,\,480$GeV$^{2}$, respectively.

\item[Fig.3]
- The function $f(x)=4[1-J_{0}(x)]/x^{2}$.

\item[Fig.4]
- The height of the plateau $H_{T,L}(Q^{2})$ and the $Q^{2}$-rescaling
factor $C_{T,L}(Q^{2})$ as a function of $Q^{2}$.
\end{itemize}
\end{document}